\documentclass[12pt]{iopart}

\usepackage{graphicx}
\usepackage{breqn}

\begin{document}

\title{Theoretical analysis of $\beta^-$ emission from $^{63}Ni$ nano-particles in glassy $^{15}P$  }

\author{S. R. Mirfayzi}

\address{Centre for Plasma Physics, School of Mathematics and Physics, Queen's University Belfast, Belfast, BT71NN}

\ead{smirfayzi01@qub.ac.uk}

\begin{abstract}
    The energy loss of $\beta^-$ emission emitting from a $^{63}Ni$ source in a phosphorus $^{15}P$ scintillation medium is theoretically studied. It has shown the $\beta$ energy spectrum absorption in ${15}P$ had nearly 100$\%$ efficiency for $\leq$ 28 keV in 800 $\mu m$ scintillator thickness. This can eventually lead to the production of light sources using these beta-emitting radiation sources as a low energy source in the near future.  
	
\end{abstract}

\section{Introduction}
The development in nanotechnology has shed a new light in production scintillating devices capable of producing lights at different wavelengths. The scintillation light generated in this method can be used to produce electric current directly using suitable converters. For instance, transparent scintillation medium such as phosphorus ($^{15}P$) can be used to create optical light via low energy $\beta^-$-emission (17 keV - 67 keV in $^{63}Ni$). Consequently, this emission can be converted to produce few volts at 100s of mA useful for micro-electro-mechanical (MEM) sensors, or be implemented in low energy boards for space and defence applications such as SiliconLab EFM low energy devices\cite{EFM32}.

However, if above applications are of interests, a thorough investigation of the emission properties, i.e. energy loss, absorption, etc., is required. Therefore, in this paper, the scattering properties and energy loss/absorption of $^{63}Ni$ source $\beta^-$-emission in a glassy $^{15}P$ scintillator nano-housing are studied. The energy loss leading to the scintillation is investigated using Bethe formula, and the cross-section is evaluated analytically for the inner-shell ionisation using Gyrzinski's method. Finally the Inelastic Mean Free Path (IMFP) is obtained and reported.  

\section{Analytical approach to emission cross-section in nano-region }
The behaviour of each particle in the scintillation medium is analytically studied in terms of transmission from initial states to final state (Bethe theory) and the energy loss to the inner shell electron excitations (Gyrzinski's method). 
The general term to determine the scattering is given by differential cross-section of an incident electron over a solid angle  $\Omega$:
\begin{equation}
d\sigma /d\Omega=\left | f^{2} \right |
\end{equation}
Where $f$ is the atomic scattering factor which is given in function of scattering angle . The differential cross-section in elastic form is represented by:
\begin{equation}
\frac{4}{a_0^{2} q^{4}} \left | f(q) \right |^{2} = \frac{4}{a_0^{2} q^{2}} \left | Z-f_x (q)) \right |^{2}
\end{equation}
where $a_{0}$ is the Bohr radius $0.529E^{-10}$ $m$, $Z$ is the atomic density and the $f_{x}(q)$ is the scattering factor for an incident electrons. This is set to zero in classical and wave mechanical theory. By having the relativistic factor $\gamma$ equals to $ 1+E_{0}/(m_{0}c^{2}) $ thus we can have:
\begin{equation}
\label{eq03}
\frac {d\sigma } {d\Omega} =\frac{4 \gamma^{2} Z^{2}} {a_0 ^{2} q^{4}}\left \{ 1- \frac{1}{(1+(qr_0)^2)^{2}} \right \}
\end{equation}
\label{eq3}
where the $q=2 k_{0} sin(\frac{\theta}{2} )$ is the momentum transfer; and $k$ is the wave vector of the electron before and after the scattering event.
\begin{figure}
	\begin{center}
		\includegraphics[width=0.8\textwidth]{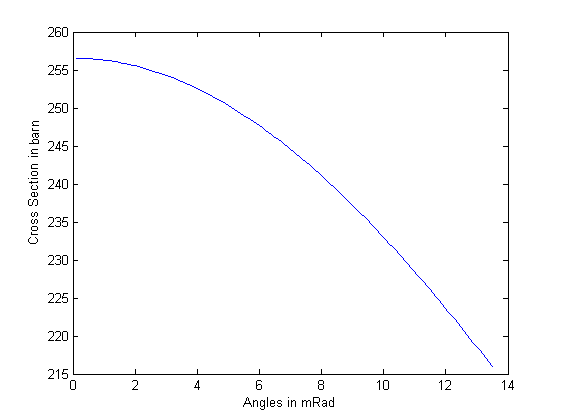}
		\caption[The Angular Dependence of Differential cross-section of 200eV in $^{15} P$]{The angular dependence of differential cross-section of 200eV in $^{15} P$. The inelastic scattering cross-section decreases as the scattering angle increases}
		\label{atomicmodel}
	\end{center}
\end{figure}
\begin{equation}
\label{eq4}
\frac{d\sigma_{i}}{d\Omega}=\frac{4\gamma^{2}Z}{a_{0}^{2}q^{4}}
\end{equation}
Where $r_{0}$ is the screening radius given by Wenz formula \cite{Wentzel} in which shows the potential nuclear attenuation as the function of $r$ distance is given by:
\begin{equation}
\label{eq5}
\phi \left ( r \right ) = \left [ Ze/4\pi \varepsilon_{0} r \right ]exp \left (\frac{-r}{r_0} \right ) 
\end{equation}
The $q$ in form of inelastic scattering will be slightly modified, since it depends on the initial and final scattering angle. Therefore in inelastic scattering it equals to $q^{2}=k_{0}^{2}(\theta ^{2}-\theta _{E}^{2})$. Here $k=\frac{2 \pi}{ \lambda} $ is the characteristic angle corresponding to average energy loss. By having the latter in an stationary state the 
$\frac{4 \gamma ^{2} Z} {a_{0}^{2} q^{2}}$ in Eq.\ref{eq3} represents the Rutherford scattering cross-section in an elastic form \cite{Schnatterly}. Using Eq.\ref{eq4} and Eq.\ref{eq5}, figure \ref{atomicmodel} is generated to demonstrate the inelastic cross-section as function of angle. 

\begin{figure}[!htbp]
	\begin{center}\label{Gyrzinski}
		\includegraphics[width=0.8\textwidth]{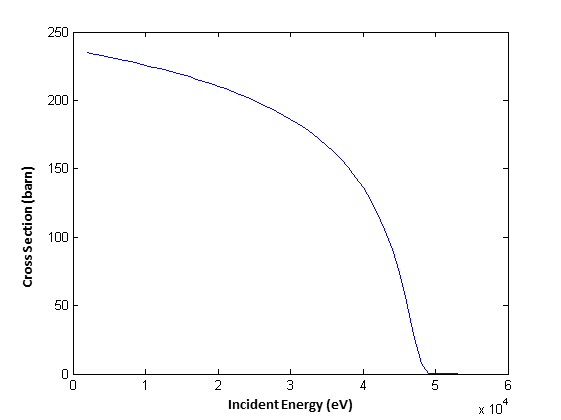}
		\caption[Gyrzinski's differential cross-section]{Gyrzinski's differential cross-section.}
	\end{center}
\end{figure}

As Figure \ref{atomicmodel} demonstrates the inelastic scattering cross-section decreases as the scattering angle increases, which means most of the scattering happens in the range of $\theta_{E}<\theta<\theta_{0}$. This is roughly proportional to $1/ \theta^{2}$ meaning a higher probability of scattering in the forward direction. 
Fig.\ref{Gyrzinski} demonstrates the differential cross-section calculated with Gyrzinski's equation. It is clear that the inner shell electrons contribute relatively little to the cross-section.
%%%%%%%%%%%%%%%%%%%%%%%%%%%%%%%%%
On the other hand, the Gyrzinski's approach enabled the calculation of the energy loss due to the band excitation. Hence the direct simulation of individual excitation specifically in the valence band will be more feasible. Since there are a limited number of electrons available in the inner shell, the inelastic stopping power is calculated by taking away the portion of energy lost to the valence band. Thus we have:
\begin{equation}
(\frac{dE}{dx})=(\frac{dE}{dx})_{total} -\Sigma_i (\frac{dE}{dx})_i
\end{equation}
Here the first term is the total energy loss and the second term representing the energy lost to valence electrons. Thus $i$ shows the number valance electron which the equation is solved for. Now the stopping power can be calculated from (for derivation see \cite{tung}):
\begin{equation}
(\frac{dE}{dx})=n_i\pi e^{4} \frac{1}{E} (\frac{E-E_i}{E+E_i})^{1.5} [ln \frac{E}{E_i}+\frac{4}{3}ln[2.7+ \frac{E}{E_i}-1]]^{0.5}
\end{equation}

Where $n_i$ is the number of electrons, $E_i$ is the binding energy, and $E$ is the incident energy. The second term demonstrates the relativistic factor, and the number 2.7 corresponds to atomic ionization potential. Thus the formula for excitation energy is given by:
\begin{dmath}
\frac{d \sigma_i}{d (\Delta E)}=n_i \pi e^{4} \frac{1}{(\Delta E)^3}\\ \frac{E_i}{E}(\frac{E}{E+E_i})^{1.5} (1- \frac{\Delta E}{E})^{E_i+ \Delta E} \times [\frac{\Delta E}{E_i}(1- \frac{E_i}{E})+ \frac{4}{3}ln[2.7+(\frac{E- \Delta E}{E_i})^{0.5}]   ]
\label{excitationeq1}
\end{dmath}
$ \Delta E $ is the energy loss. Now cross-section can be calculated from:
\begin{dmath}
\label{emc}
\sigma_i= n_i \pi e^{4} \frac{1}{E_i ^{2}} \frac{E_i}{E}(\frac{E-E_i}{E+E_i})^{1.5}[1+ \frac{2}{3}(1- \frac{E_i}{2E}) \times ln[2.7+(\frac{E}{E_i}-1)^{0.5}]]
\label{excitationeq2}
\end{dmath}
Girzinsky’s formula provides an easy approach to calculate the relative energy loss. The simulation is done for $200eV$ incident electron interacting with three inner shell electrons of $^{15}P$. Using equation \ref{excitationeq1} and \ref{excitationeq2} from the incoming 200 eV, the energy of 69.6 eV was spent on excitation energy only. 

Now that the loss due to excitation is determined using Gyrzinski’s approach we now can verify the loss using Bethe formula. The Bethe theory can be used to evaluate the behaviour of each atomic electrons in terms of the transition from initial state to final state. 
This can be approached using first Born approximation \cite{Inokuti}; therefore, the differential cross-section can be determined by:
\begin{dmath}
d \sigma_n =(2 \pi)^{-2} M^{2}\hbar ^{-4}(k^{'}/k)\int exp(iK.r)u_n \ast (r_1,..., r_z) \times V u_0(r_1,..., r_z) dr_1...dr_z dr)|^{2}dw,
\end{dmath}
where $M$ is the reduced mass of colliding system, $r$ is the position of particle relative to the centre of atom, $\hbar k$ is the momentum of particle before the collision and $\hbar^{‘}$ is the momentum after the collision, $u$ is the eigenfunction of $r_j$ coordinates whose the total number is $Z$. Now we can have the Eq.$2.10$ in angular form:
\begin{equation}
\frac{d \sigma_n}{d \Omega}=( \frac{m_0}{2 \pi \hbar^{2}})|\int V(r) \psi_0 \psi_n^{*} exp(iq.r)d\tau  |^{2}
\end{equation}
Here the initial states and final states are $\psi_0$ and $ \psi_n^{*}$. $k_0$ and $K_1$ are the wave-vectors and $V(r)$ is the potential energy required by each collision.  The $V(r)$ is required due to existence of Coulomb force, and can be calculated using:
\begin{equation}
V(r)= \frac{Ze^{2}}{4 \pi \varepsilon_0 r }- \frac{1}{4 \pi \varepsilon _0} \sum_{j=1}^{Z} \frac{e^{2}}{|r-r_j|}
\end{equation}

Where the first term represents the Rutherford scattering and $|\varepsilon _n (q)|^{2}$ is the dynamical structure factor. The dynamical structure factor is closely related to the generalised oscillator strength \cite{Inokuti}, is calculated using Rydberg energy $(13.61eV)$ and transition energy. 
%%%%%%%%%%%%%%%%%%%%%%%%%%%%%%%%%%%%%
\begin{figure}[!htbp]
	\begin{center}
		\leavevmode
		\includegraphics[width=0.8\textwidth]{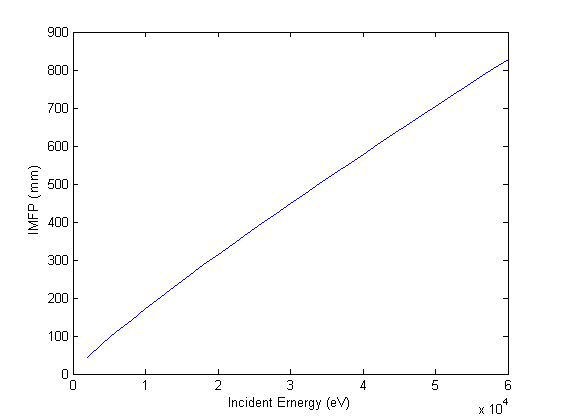}
		\caption[Inelastic Mean Free Path in $^{15}P$]{Inelastic Mean Free Path in $^{15}P$.}
		\label{imfp}
	\end{center}
\end{figure}
The loss due to ionization and excitation is measured the Inelastic Mean Free Path (IMFP) and the model describing IMFP is based on a theoretical approach to determine the transport of primary electron beams on the surface. In this matter, a relationship between signal intensity and concentration of given elements in an elastic and inelastic deferential cross-section is required. This could be a complex function as the accurate data from elastic and inelastic cross-section is not obtainable. Hence the computational functions are usually solved by looking into continuous slowing down approximation (CSDA) in which an electron energy along the trajectory is assumed to be a function of length. To tackle this problem the excitation function for a charged particle is defined by its dielectric function as demonstrated by \cite{Pines}:

\begin{equation}
\frac {d^{2} \sigma}{d (\Delta E) dq}=\frac{m e^{2}}{\pi \hbar^{2} NE } Im(\frac{-1}{\varepsilon (\omega , q)}) \frac{1}{q}
\end{equation}
Where $N$ is number density of the atom, $E$ is the energy of incident beam, $(\Delta E) $ is the energy loss and $ Im(\frac{-1}{\varepsilon (\omega , q)})$ is the dielectric function. The dielectric function can be reduced to a more generalised form of Lindhard type dielectric function \cite{Ganachaud} and \cite{Desalvo}. In This model the IMFP is calculated based on \cite{Penn} and \cite{Tanuma}approach:
\begin{dmath}
\lambda = \frac{E}{E_p^2[\beta ln( 0.191 \rho^{-0.5}E)-(1.97-0.91(N_v \rho /M)/E)+((53.4-20.8(N_v \rho /M)/E^{2})]}
\end{dmath}
Here the $\rho$ is the material density and $E_g$ is the band gap energy, $M$ is the atomic weight, $\beta = -0.1+0.944(E_p^{2}+E_g^{2})^{-0.5}+0.069 \rho^{0.1})$, $E_P=28.8(N_v \rho /M)^{0.5}$ is the free electron plasmon energy, and $N_v$ is the number of valence electron per atom. 

Figure \ref{imfp} demonstrates the evaluated IMFP for $^{15}P$. In this case the $N_v$ is set to $5$, $M=30.97$, $E_q=3eV$ and $\rho=1.88 g/cm^{3}$. These are the value which are typical for our material (data from NIST library).

\section{The energy loss analysis of the ${63}Ni$ $\beta^{-}$ energy spectrum in ${15}^P$} 

\begin{figure}[!htbp]
	\begin{center}
		\includegraphics[width=0.8\textwidth]{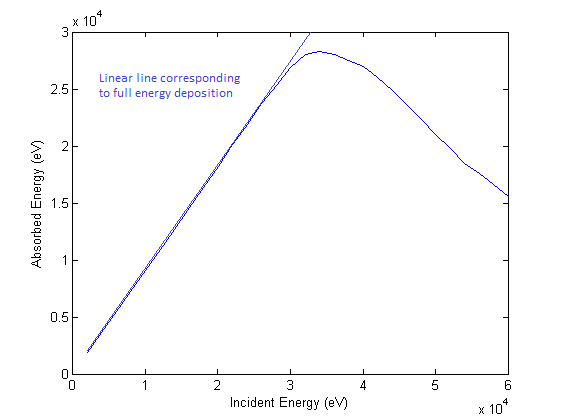}
		\caption[The Energy Absorbed in $^{15}P$]{The Energy Absorbed in $^{15}P$. The film of $^{15}P$ is set to have 800 $\mu m$. A monoenergetic particle source is created with different energies between $E_{min} =2keV$ and $E_{max}=60keV$ to simulate the $^{63} Ni$ spectrum. As figure denotes, at energies over 28keV the incident particles starts to deposit most of their energies deeper in the film with some also penetrating, this means the IMFP is increases. This also corresponds to higher particle range}
		\label{energylossatphos}
	\end{center}
\end{figure}

The main process of absorption is the excitation, ionization. As Fig.\ref{energylossatphos} shows, at energies over $28$keV the incident particles start to deposit most of their energies in the film (800 $\mu m$ thick) and starts escaping at higher energies, this means the IMFP is increased. 

The energy loss in 2 keV to 28 keV range is almost 99.9$\%$. Above 28keV a portion of incoming particles escapes the film. Fig.\ref{energylossatphos} represents the ${63}Ni$ $\beta^{-}$ energy spectrum absorption in ${15}^P$, and the linear line corresponding to full energy deposition (100$\%$ efficiency)
Particle range consists of four profiles, longitudinal range, lateral straggling, projected range, and collisional range. The longitudinal range increases because of increase in energy spread (of the incoming beam), which is not influenced by dispersion effect. This has a direct relation to beam profile. Although this could be sometimes negligible particularity for a single interaction, however, it cannot be ignored for a continuous beam. Considering the initial beam profile be $ \varepsilon _z$ the final Root Mean Square (RMS) is given by \cite{Thangaraj}:   
\begin{equation}
\label{emc2}
\varepsilon^2_{z final}= \varepsilon_z^2+(\frac{17 \lambda^2}{40D})^2 \left \langle X^2 \right \rangle[\left \langle Z^2 \right \rangle + \alpha^2 D^2 \left \langle \delta ^2 \right \rangle + 2\alpha D \left \langle Z \delta \right \rangle]
\end{equation}
Where $ \alpha D$ is the longitude dispersion, D is the dispersion factor and $X’, Z, \delta $ are the angle, position and energy spread respectively. More studies can be done during the experimental prototyping and the effect of beam dispersion could be taken into account before a valid comparison between the final prototype and simulation can be made.

Another dependent of particle range is lateral scattering which is significantly affected by the size of the incoming beam, and this is because of correlations between the particle range and its position. Finally, the projected mean range is calculated from CSDA function. The simplified mean free particle range hence is given by \cite{Rose}:
\begin{equation}
\left \langle R \right \rangle = t(1+t/ \lambda_0)
\end{equation}
Where $\lambda_0$ is mean free path of electrons and limited to the value of thickness (t). When t is approaching R (range), the corresponding value of R can be called L (mean free range), thus L can be recalled as:
\begin{equation}
R=\lambda \frac{ \left [ (4L/ \lambda)+1 \right ]^2 -1 }{2}
\end{equation}
Here $\lambda= \lambda_0 /4$ and can be given as \cite{tabata}:
\begin{equation}
\lambda_0=2 a_0^{-1} (Z) E_0^{\alpha_n(z)}
\end{equation}
In an experimental approximation $\alpha_n(z)$ is the tabulated thickness where $n=1,2,...$. This leads to the definition of collisional range is where the incoming beam deposits most of their energy due to the collision. This value represents an accumulated depth in the film where most of the interactions take place. This could later facilitate an understanding of an optimum thickness required for the application. 

\section{Conclusion}
This report has discussed modelling of  $^{63}Ni$ $\beta$-source doped in phosphorous ($^{15}P$). The energy loss processes are discussed and analysed theoretically. 
%\section{Acknowledgement}

\section*{References}
% ------------------------------------------------------------------------
%  BIBLIOGRAPHY FILE
% ------------------------------------------------------------------------
%\bibliography{iopart-num}

\begin{thebibliography}{10}
	\expandafter\ifx\csname url\endcsname\relax
	\def\url#1{{\tt #1}}\fi
	\expandafter\ifx\csname urlprefix\endcsname\relax\def\urlprefix{URL }\fi
	\providecommand{\eprint}[2][]{\url{#2}}
	\bibitem{EFM32} EFM32, https://www.silabs.com/
	
	\bibitem{franklin}
	Franklin P 1958 {\em An Introduction to Fourier Methods and the Laplace
		Transformation\/} (New York: Springer) p 158
	
	\bibitem{widlund}
	Widlund O~B 1987 {\em International Symposium on Domain Decomposition Methods
		for Partial Differential Equations\/}  113--128
	
	\bibitem{Babuska}
	Babuska I and Gatica G~N 2003 {\em Nubmers and Methods Partial Differential
		Equations\/}  192--210
	
	\bibitem{GeigerMarsden}
	Geiger C and Marsden E 1909 {\em Proc. R. Soc. Lond\/} {\bf A82} 495--500
	
	\bibitem{ReimerKohl}
	Reimer L K~H 2008 {\em Transmission Electron Microscopy: Physics of Image
		Formation\/} (5th edition, Springer)
	
	\bibitem{Wentzel}
	Wentzel M~R and Vos M 2006 {\em Nucl. Instrum. Methods Phys\/} {\bf Res. B}
	998--1011
	
	\bibitem{Schnatterly}
	Schnatterly S~E 1979 {\em Solid State Physics\/} {\bf 14} 275--358
	
	\bibitem{Ding}
	Ding Z~J and Shimzu R 1988 {\em Surface Science\/} {\bf 197} 539--554
	
	\bibitem{tung}
	CJ~Tung J~A and Ritchie R 1979 {\em Surface Science\/} {\bf 81} 427
	
	\bibitem{Inokuti}
	Inokuti M 1971 {\em Rev. Mod. Phys.\/} {\bf 43} 997--347
	
	\bibitem{Banhart}
	FBanhart 1999 {\em Rep.Prog.Phys.\/} {\bf 62} 1181
	
	\bibitem{Pines}
	Pines D and Nozieres P 1966 {\em Rep.Prog.Phys.\/} {\bf 1}
	
	\bibitem{Ganachaud}
	Ganachaud J and Cailler M 1979 {\em Surface Sci.\/} {\bf 83} 498
	
	\bibitem{Desalvo}
	A~Desalvo A~P and Rosa P 1984 {\em J. Phys.\/} {\bf D17} 2455
	
	\bibitem{Penn}
	Penn D~R 1978 {\em Physical Review B\/} {\bf 35} 483--486
	
	\bibitem{Tanuma}
	S~Tanuma C~J~P and Penn D~R 1991 {\em Surf. and Interface Anal.\/} {\bf 17} 927
	
	\bibitem{Thangaraj}
	J~Thangaraj J~Ruan A~S~J~R~T~K~A~H~L~J~S~Y~E~S~T~M~H~E 2012 {\em IEEE,
		Proceedings of IPAC2012\/}  49--51
	
	\bibitem{ICRU}
	ICRU 1972 {\em International Commission on Radiation Units and Measurements
		Report\/} {\bf 21}
	
	\bibitem{Rose}
	Rose M~E 1940 {\em Phys. Rev.\/} {\bf 58}(1) 90--90
	
	\bibitem{tabata}
	Tabata T 1968 {\em Journal of Applied Physics\/} {\bf 39} 5342--5343
	
\end{thebibliography}
\providecommand{\newblock}{}

% ------------------------------------------------------------------------
% ------------------------------------------------------------------------

\end{document}